\documentclass[aps,twocolumn,prb,floatfix,showpacs,superscriptaddress]{revtex4-1}

\usepackage{amsmath}
\usepackage{amssymb}
\usepackage{amsfonts}
\usepackage[pdftex]{graphicx}
\usepackage{units}
\usepackage[usenames]{color}
\usepackage{dcolumn}
\usepackage{bm}
\usepackage{float}
\usepackage[utf8]{inputenc}
\usepackage[colorlinks=true,citecolor=blue,linkcolor=blue]{hyperref}

\usepackage{epstopdf}

\definecolor{cream}{RGB}{222,217,201}
\definecolor{orange}{RGB}{255,128,0}
\definecolor{teal}{RGB}{0,128,128}
\definecolor{violet}{RGB}{127,0,255}

\begin{document}

\title{High-throughput computational discovery of CuI-based ternary $p$-type transparent conductors}

\author{Michael Seifert}
\affiliation{Research Center for Future Energy Materials and Systems and
  Interdisciplinary Centre for Advanced Materials Simulation,
  Ruhr University Bochum, Universit\"atsstra{\ss}e 150, D-44801 Bochum, Germany}
\affiliation{ Faculty of Physics and Astronomy, Ruhr University Bochum,
  Universit\"atsstra{\ss}e 150, D-44801 Bochum, Germany}
\affiliation{Institut f\"ur Festk\"orpertheorie und -optik,
  Friedrich-Schiller-Universit\"at Jena, Max-Wien-Platz 1, 07743 Jena,
  Germany}
\affiliation{European Theoretical Spectroscopy Facility}
\author{Miguel A. L. Marques}
\affiliation{ Faculty of Physics and Astronomy, Ruhr University Bochum,
  Universit\"atsstra{\ss}e 150, D-44801 Bochum, Germany}
\affiliation{European Theoretical Spectroscopy Facility}
\author{Silvana Botti}
\affiliation{Research Center for Future Energy Materials and Systems and
  Interdisciplinary Centre for Advanced Materials Simulation,
  Ruhr University Bochum, Universit\"atsstra{\ss}e 150, D-44801 Bochum, Germany}
\affiliation{ Faculty of Physics and Astronomy, Ruhr University Bochum,
  Universit\"atsstra{\ss}e 150, D-44801 Bochum, Germany}
\affiliation{Institut f\"ur Festk\"orpertheorie und -optik,
  Friedrich-Schiller-Universit\"at Jena, Max-Wien-Platz 1, 07743 Jena,
  Germany}
\affiliation{European Theoretical Spectroscopy Facility}

\begin{abstract}
P-type transparent conducting materials (TCMs) remain scarce, limiting
progress in transparent electronics and tandem photovoltaics. Here, we
perform a high-throughput computational search across CuI-based ternary
X--Cu--I compounds using the minima hopping method combined with density
functional theory. Filtering for thermodynamic stability, optical
transparency, and light hole effective mass, we identify 58 candidate
p-type TCMs, of which 49 are previously unreported. Twenty are p-type
degenerate semiconductors with intrinsic hole carriers, including
Cu$_2$FI$_2$ (calculated gap 3.75~eV, hole effective mass $0.262\,m_0$).
We further use this dataset to revisit the chemical modulation of the
valence band (CMVB) design rules, originally formulated for Cu-based
delafossite oxides. Four generalizations emerge: linear Cu coordination is not required, as tetrahedral coordination (CN~=~4) is compatible with p-type TCM character; the mechanism extends to halide-based systems with I~$5p$ anion states; the
Cu $d$ state energy center $\epsilon_d$ is a quantitative descriptor for
valence band dispersion; and Cu undercoordination in monovalent-anion
hosts drives intrinsic p-type carrier generation through stoichiometric
charge balance. These results establish a generalized computational
framework for the design of p-type halide TCMs.
\end{abstract}

\maketitle

\section{Introduction}

\label{sec:introduction}

The concept of transparent conducting devices has been discussed for
several decades, driven by the prospect of merging electrical
conductivity with optical transparency in the visible spectral range.
Such a combination would enable a new generation of functional electronic
devices, including transparent solar
cells~\cite{Husain_2018, AoLiu_2021, ChenchenYang_2019} that can be
integrated into building facades, transparent
displays~\cite{AoLiu_2021, Willis_2021} on windows, transparent heating
elements~\cite{Chopra_1983, Yang_2017, Gupta_2016}, and transparent
transistors. Realizing these applications requires suitable transparent
conducting materials (TCMs) for both carrier types. While n-type TCMs such
as ITO~\cite{Minami_2008}, ZnO~\cite{Liu_2013}, and
SnO$_2$~\cite{Afre_2018} are well established in industry, finding
high-performance p-type counterparts has remained a persistent challenge.
Oxide-based p-type candidates, which have been studied extensively,
generally suffer from insufficient hole
mobility~\cite{Grundmann_2013, Willis_2021, Fioretti_2020, Hu_2020,
AoLiu_2020}.

The conceptual foundation for the rational design of p-type transparent
conducting oxides was established by Kawazoe and co-workers, who
demonstrated p-type conductivity in transparent thin films of the
delafossite CuAlO$_2$~\cite{Kawazoe_1997}. The key insight, formalized
as the ``chemical modulation of the valence band'' (CMVB), is that
hybridization of closed-shell Cu$^+$ 3$d^{10}$ states with the anion
$p$ states at the valence band maximum (VBM) raises the VBM energy,
delocalizes hole states, and leads to a more dispersive valence band
with lower hole effective mass. This design principle was further
developed and applied to a broader family of Cu-based delafossites and
related materials~\cite{Kawazoe_2000, Kawazoe_1997, Yanagi_2000}.

\begin{figure}[htpb!]
\centering
\includegraphics[width=9cm]{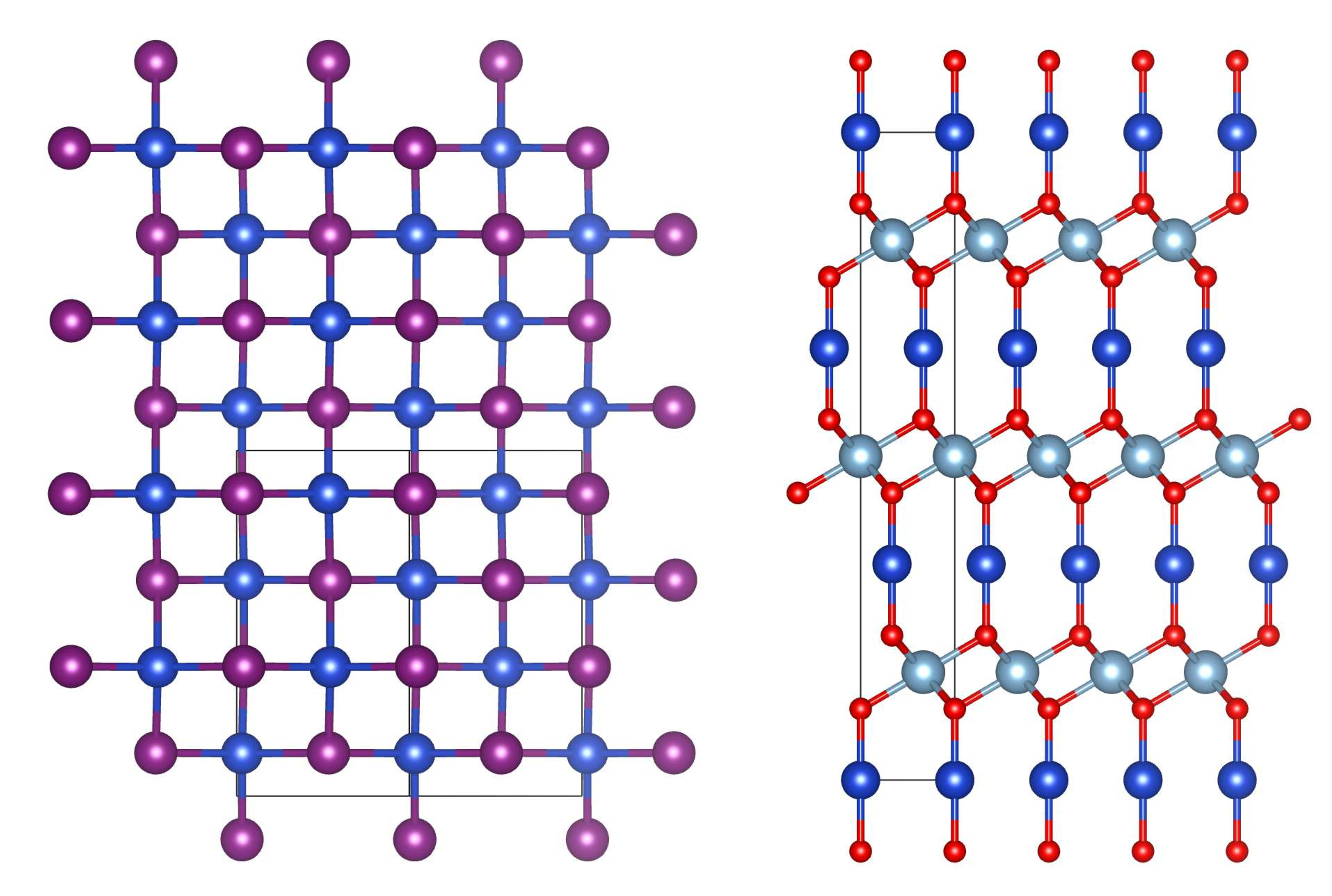}
\caption{Crystal structures of CuI (left) and CuAlO$_2$ (right) as representative p-type TCMs. Dark blue balls represent Cu, violet balls I, red balls O, and light blue balls Al. The figure was made with VESTA~\cite{momma_izumi_2011}.}
\label{Fig_StructPTCM}
\end{figure}

As an alternative to oxide-based p-type TCMs, cuprous iodide (CuI) has
experienced a revival of interest after its initial discovery in the
early twentieth century~\cite{Baedeker_1909, Steinberg_1911}. At room
temperature, CuI crystallizes in the zincblende structure (referred to
as $\gamma$-CuI), and combines a wide band gap of
3.1~eV~\cite{Krueger_2018} with a high hole mobility of
40~cm$^2$V$^{-1}$s$^{-1}$~\cite{Yang_2017} and a carrier density in
the range 10$^{16}$--10$^{19}$~cm$^{-3}$~\cite{Chen_2010, Yang_2016},
arising naturally from copper vacancies as intrinsic
defects~\cite{Wang_2011, Huang_2012, Jaschik_2019, Grauzinyte_2019}.
A large exciton binding energy of approximately
60~meV~\cite{Grundmann_2013, Krueger_2021} points to further potential
for optoelectronic applications, and a high thermoelectric figure of
merit has also been reported~\cite{Yang_2017, Grundmann_2013}. CuI has
accordingly been incorporated in a variety of devices, including solar
cells~\cite{Christians_2014, AoLiu_2021},
photodetectors~\cite{Yamada_2019}, LEDs~\cite{Ahn_2016, Baek_2020},
transistors~\cite{Choi_2016, Tixier_2016}, and thermoelectric
devices~\cite{Yang_2017, Almasoudi_2022}.

Despite these promising properties, CuI still lags behind established
n-type TCMs in hole mobility~\cite{Yang_2016room}, and the modulation of
carrier density for active device design has proven
difficult~\cite{AoLiu_2021}. To address these shortcomings, doping and
alloying strategies have been actively explored in recent years.
Regarding doping, chalcogen elements have shown particular promise as
acceptors~\cite{Storm_2021_Se, Ahn_2022}, and a systematic
computational study of 64 substitutional impurities in $\gamma$-CuI
identified chalcogens on the iodine site as the most effective p-type
dopants, with eight elements also viable for n-type
doping~\cite{Grauzinyte_2019}. For alloying, both cation and anion
substitution have been investigated. On the cation side,
Ag~\cite{Annadi_2020_AgCuI, Krueger_2023_AgCuI},
Ni~\cite{Annadi_2020_NiCuI, Dethloff_2024}, and
Zn~\cite{Yamada_2022} have been studied. On the anion side,
substitution by other halogen
atoms~\cite{Yamada_2020, Seifert_2022_CuBrI} and by chalcogen
elements~\cite{Seifert_2024} has demonstrated tunability of the optical
band gap, the carrier density, and the band topology of CuI-based
ternary compounds.

The CMVB design rules, originally formulated for Cu-based delafossite
oxides with linearly coordinated Cu$^+$ in an O--Cu--O dumbbell motif,
have not been systematically tested across a broad and structurally
diverse set of CuI-based non-oxide compounds. In particular, it is an
open question whether the key ingredients identified by Kawazoe and
co-workers, such as $d^{10}$ cation character, $p$-$d$ hybridization
at the VBM, and an appropriate coordination environment for Cu, are
necessary, sufficient, or generalizable to the broader class of
halide-based p-type TCMs where Cu is predominantly tetrahedrally
coordinated. The structural diversity of predicted CuI-based ternary
candidates, encompassing zincblende-derived, layered, and low-symmetry
phases, is illustrated in Fig.~S2 and Table~S4 of the Supporting Information.

To address this, we present an unbiased computational
search for p-type TCMs among CuI-based ternary compounds of the form
X--Cu--I, where X spans a wide selection of elements from the periodic
table, chosen consistently with earlier high-throughput materials
searches~\cite{Cerqueira_2015, Koerbel_2016}. The corresponding ternary
phase diagrams were explored using the minima hopping method
(MHM)~\cite{Amsler_2010, Goedecker_2004} combined with density
functional theory (DFT). After applying filtering criteria based on
thermodynamic stability, band gap, and hole effective mass, 58 candidate
p-type TCMs are identified. To our knowledge, all but three are unreported
in the literature, beyond the sulfur- and selenium-based ternaries we
characterized in our earlier work~\cite{Seifert_2024}. This dataset is
then used to systematically test each of the CMVB design rules against a
structurally diverse, non-oxide family of CuI-based compounds, assessing
which rules transfer, which require generalization, and which new
descriptors emerge from the data. In doing so, we achieve a substantial
extension of the current knowledge of CuI-based ternary compounds as
candidate p-type TCMs, and provide a computational foundation for guiding
future experimental synthesis efforts.

The manuscript is organized as follows. Section~\ref{sec:methods}
describes the computational methods, covering the structure prediction
procedure, the thermodynamic stability analysis, and the electronic
characterization steps. Section~\ref{sec:results} presents the results
of the high-throughput filtering. Section~\ref{sec:design} constitutes
the core of the paper: taking the 58 candidate materials as a dataset,
we address each CMVB design rule in turn, characterizing the bonding,
coordination environment, $p$-$d$ hybridization, valence band
dispersion, and carrier generation across the set.
Section~\ref{sec:conclusion} summarizes the main findings and outlines
directions for future work.

\section{Computational Details}

\label{sec:methods}

\begin{figure}[htpb!]
\centering
\includegraphics[width=9cm]{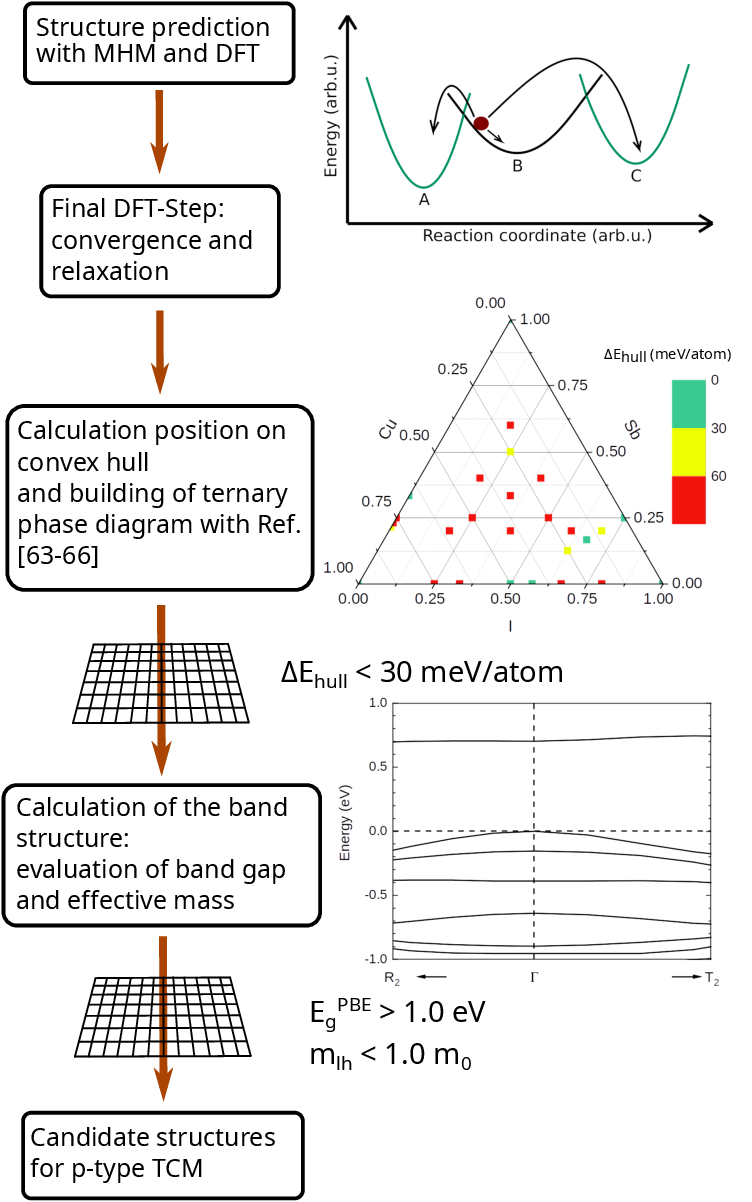}
\caption{Workflow of the high-throughput study for predicting ternary
  CuI-based compounds as candidate p-type TCMs. The three sequential
  filter steps address thermodynamic stability, optical transparency,
  and hole mobility.}
\label{Fig_Workflow}
\end{figure}

\subsection{Structure prediction}

The prediction of stable crystalline phases requires an exploration of
configuration space that encompasses, in principle, all crystal
structures at all possible stoichiometries for the elements under
consideration. In the case of ternary systems, this means exploring the
full set of stable ternary, binary, and elemental phases. At zero
temperature and pressure, the convex hull of thermodynamically stable
structures defines the set of low-energy phases that cannot decompose
into other compounds. To identify these phases, we employed the minima
hopping method (MHM)~\cite{Amsler_2010, Goedecker_2004}, a structure
prediction algorithm designed to locate the most favorable crystal
structure for a given stoichiometry. The MHM evaluates the enthalpy
hypersurface at fixed pressure through consecutive short molecular
dynamics steps and local geometry relaxations of both atomic positions
and cell shapes. Initial atomic velocities are approximated along
soft-mode directions, enabling efficient escapes from local minima
toward lower-energy configurations. A feedback mechanism prevents
revisiting of already-found minima. The MHM has been successfully
applied in numerous
contexts~\cite{Huan_2013, Shi_2017, Borlido_2020, SunLin_NatComm_2021,
SunLin_2022}, including the Cu--I binary
system~\cite{Jaschik_2019}.

Total energies and forces for the MHM were computed using density
functional theory (DFT) with the projector augmented wave (PAW)
method~\cite{Bloechl_1994}, as implemented in
VASP~\cite{Kresse_1996, kresse_joubert_1999}. PAW potentials were
employed consistently with the Materials Project~\cite{MP}. In the
structure prediction runs, default plane-wave cutoff energies were
applied together with a $\mathbf{k}$-point density of
0.025~\AA$^{-1}$ in reciprocal space, using $\Gamma$-centered
Monkhorst--Pack grids. The Perdew--Burke--Ernzerhof (PBE) generalized
gradient approximation~\cite{Perdew_1996} was used for the
exchange-correlation functional throughout the structure prediction
stage. For transition metals acting as the third element X, spin
polarization was omitted during the initial structure prediction stage.
Ni-containing candidates that survived the filtering procedure were
subsequently evaluated in their spin-polarized ground states.
Consistent with experimental observations~\cite{Dethloff_2024}, these
calculations confirmed ferromagnetic behavior for the Ni--Cu--I
compounds.

MHM runs were initialized from single random configurations to avoid
starting-configuration bias. Initial atomic positions were constrained
such that interatomic distances exceeded the sum of the respective
covalent radii, preventing unphysical geometries. For each ternary
phase diagram, a full set of stoichiometries was sampled: all
compositions with up to 6 atoms per unit cell were considered
systematically, and if stable binary phases were known, the
corresponding ternary composition formed as their sum was additionally
included. The complete set of ternary phase diagrams is provided in the
Supporting Information. Each MHM run was terminated after 70 distinct
local minima had been identified per stoichiometry. We note that this
procedure does not guarantee that the global minimum has been found.
Two sources of incompleteness should be kept in mind: larger unit cells
of the same stoichiometry may yield lower formation energies per atom
due to structural distortions, and entropic contributions, neglected
here, can stabilize disordered phases relative to their ordered
counterparts at finite temperature. Nevertheless, the ability of the
MHM to predict new structures, as demonstrated by previous applications
including the ferroelectric perovskite
LaWN$_3$~\cite{Talley_2021, Sarmiento-Perez_2015}, supports the
validity of the chosen parameters.

\subsection{Assessment of thermodynamic stability}

The comparatively coarse DFT parameters used during the MHM search are
necessary given the large number of energy and force evaluations required to
explore configuration space. All structures found by the MHM were
therefore recalculated in a final high-accuracy DFT step to determine
their positions on the convex hull (see Fig.~\ref{Fig_Workflow} for an
overview of the workflow). In these refinement calculations,
$\mathbf{k}$-point grids were converged individually to a total energy
tolerance of 1~meV/atom using $\Gamma$-centered Monkhorst--Pack meshes.
The plane-wave cutoff energy was set to 700~eV for all structures. Ionic
relaxations were repeated until all residual forces were below
1~meV/\AA, and electronic self-consistency was converged to
$10^{-6}$~eV.

Convex hull construction at zero temperature and pressure required
reference energies for all competing binary and ternary phases. These
were taken from the Materials Project~\cite{MP} and supplemented with
data from our own prior
work~\cite{Schmidt_2022, Schmidt_2023, Schmidt_2024, Cavignac_2026}.
All reference structures were recalculated with our own convergence
parameters to ensure consistency. For structures within 30~meV/atom of
the PBE convex hull, corresponding approximately to the thermal energy
at room temperature and adopted as the stability threshold throughout
this work, the geometry optimization was repeated using the
PBEsol~\cite{Perdew_2008} and SCAN~\cite{SCAN} functionals, with
$\mathbf{k}$-point grids of 8000 points per reciprocal atom. The
resulting structures were added to the database of
Ref.~\cite{Schmidt_2022}. PBEsol is a variant of PBE optimized for
solid-state properties, while SCAN is a meta-GGA functional satisfying
all 17 known exact constraints on the exchange-correlation functional.
Including these two functionals alongside PBE is motivated by the known
tendency of PBE to produce significant errors in formation energies for
certain
systems~\cite{Stepanovic_2012, Perez_2015, Tran_2016, Schmidt_2022,
Bartel2019}, and their use provides additional validation of the
stability assignments for the most promising candidate structures.

\subsection{Electronic characterization}

Band gaps and hole effective masses were calculated from PBE band
structures for all structures passing the stability filter. Transparency
was assessed using the criterion $E_\mathrm{g}^\mathrm{PBE} > 1.0$~eV,
motivated by the fact that the PBE band gap of $\gamma$-CuI lies near
this value while its experimental gap is 3.1~eV~\cite{Krueger_2018}.
Structures with a smaller PBE gap than $\gamma$-CuI were discarded as
likely to absorb part of the visible spectrum; for all remaining structures,
band structures were recalculated using the modified HSE06 hybrid functional (mHSE06). Following
our earlier work on CuI-based
systems~\cite{Seifert_2024, Dethloff_2024}, the mixing parameter
$\alpha$ was set to 0.32 so that the modified functional reproduces the
experimental band gap of $\gamma$-CuI of 3.1~eV~\cite{Krueger_2018}.
An exception are CuTe$_2$I and Cu$_2$GaI$_5$. They were ultimately kept as their PBE band gap is very close to 1.0~eV and displayed large mHSE06 band gaps which are in the region of the other materials which passed this step.

Hole effective masses were evaluated by fitting even-order polynomials
up to sixth degree to the band dispersion in a window of $\pm$25~meV
around the valence band maximum (VBM), along paths connecting the VBM
to all relevant high-symmetry points in the Brillouin zone.
High-symmetry point coordinates were obtained using the SeeK-path
tool~\cite{Hinuma_2017}. Rather than computing a spatial average, we
report the minimum hole effective mass observed over all considered
paths. This choice is motivated by the large fraction of layered
structures in the dataset, for which in-plane and out-of-plane
effective masses differ by orders of magnitude and a spatial average
would be physically misleading. The adopted threshold of
$m_\mathrm{lh} < 1.0\,m_0$ is more stringent than the spatially
averaged criterion of $1.5\,m_0$ used in Ref.~\cite{CEEM_2024}.

For the characterization of $p$-$d$ hybridization, the mHSE06 functional was applied in addition to PBE.
Hybrid functionals provide a more accurate description of the
localization of Cu $d$ states than semilocal functionals, and are
therefore better suited for a quantitative assessment of $p$-$d$
hybridization. Effective masses were obtained from PBE band structures
in all cases, as PBE and hybrid functionals are expected to yield
similar valence band dispersions.

Spin-orbit coupling was omitted for the calculation of the electronic structure. For $\gamma$-CuI it is known that it will just cause a slight shift in the band gap and there is the splitting of the split-off band from the heavy and light hole band at the VBM. Similar observation were made in our past works together with experimental data regarding Cu(Br,I)~\cite{Seifert_2022_CuBrI} and (Cu,Ag)I~\cite{Krueger_2023_AgCuI}. Therefore, the general trends and observation we discuss here are expected to be valid.

\subsection{Chemical bonding analysis}

Chemical bonding analysis was performed using the LOBSTER
package~\cite{Lobster} as a post-processing step for VASP calculations.
For each structure, we computed the Integrated Crystal Orbital Bonding
Index (ICOBI)~\cite{Ertural_2021}, which provides a measure of bond
covalency: a value of 1 indicates a purely covalent bond, while lower
values reflect increasing ionicity. The Crystal Orbital Hamiltonian
Population (COHP)~\cite{Dronskowski_1993} was computed to characterize
the bonding and antibonding character of states near the VBM, as both
$p$-$d$ and $s$-$p$ hybridization are known to produce antibonding
states at the VBM that are linked to high hole
mobility~\cite{Williamson_2017, Xu_2018}.

The LOBSTER calculations used PBE wavefunctions as input. The
$\mathbf{k}$-point sampling density was increased relative to the value
used for total-energy convergence by incrementing the subdivision
parameter by 20, ensuring adequate sampling of the Brillouin zone for
the bonding analysis. Atomic basis functions were chosen to match the
orbitals treated as valence in the VASP PAW potentials. Charge spilling
was verified to remain below the threshold established in
Ref.~\cite{Naik_2023}, and results for $\gamma$-CuI were confirmed to
be consistent with the ICOBI and ICOHP values reported therein. All
nearest-neighbor interactions were included in the bonding analysis.
Nearest neighbors were identified from the VASP structural output using
Robocrys~\cite{Ganose_2019}, with visual inspection applied in cases of
ambiguity. For layered structures, only atoms within the same layer were
treated as nearest neighbors, as interlayer distances are large and
interlayer interactions are dominated by van der Waals forces, which are
not captured by the ICOBI index.

\subsection{Valence band descriptors}

Three electronic descriptors were computed across the dataset of
candidate materials to characterize the valence band and its
relationship to hole transport.

To address the question of orbital hybridization, we obtained the
projection of the wavefunction at the VBM onto $s$, $p$, and $d$
orbitals, denoted $\alpha_i$ for $i \in \{s, p, d\}$, following the
notation of Eq.~2 in Ref.~\cite{Cardona_1963}. The center of the Cu
$d$ states, $\epsilon_d$, was obtained as the energy-weighted first
moment of the $d$-projected density of states,
\begin{equation}
    \epsilon_d = \frac{\int \mathrm{DOS}(E)\, E\, \mathrm{d}E}
      {\int \mathrm{DOS}(E)\, \mathrm{d}E},
\end{equation}
integrated over the Cu $d$ states. In cases where the third element X
contributes $d$ states that mix with the Cu $d$ states near the VBM (as
occurs for X = Ni and X = Ag), the $d$ states of X were included in the
integral. When the $d$ states of X are well separated from the Cu $d$
states in energy, as for X = Ga (whose $d$ states lie near $-15$~eV),
they were excluded, as they do not contribute to the valence band
dispersion near the VBM. The center $\epsilon_d$ is referenced to the
VBM rather than the Fermi level, so that the comparison between
conventional semiconductors and p-type degenerate semiconductors (in
which the Fermi level lies below the VBM) is made on a consistent
footing (see Fig.~S1 of the Supporting Information). This descriptor is
motivated by its established role as a predictor of surface reactivity
and chemisorption properties in transition
metals~\cite{Hammer_1996, Ruban_1997, Christoffersen_2001}. Here we
explore whether it carries analogous predictive power for valence band
dispersion in p-type semiconductors.

The valence band width, $\Delta E_\mathrm{VB}$, is defined as the
maximum energy spread of the bands forming the VBM across the full
Brillouin zone, providing a measure of the overall dispersion of the
highest occupied band beyond the local vicinity of the VBM captured by
the effective mass. In the case of band degeneracies, the maximum energy
spread of the band manifold determines $\Delta E_\mathrm{VB}$.

\begin{figure*}[htpb!]
\centering
\includegraphics[width=16cm]{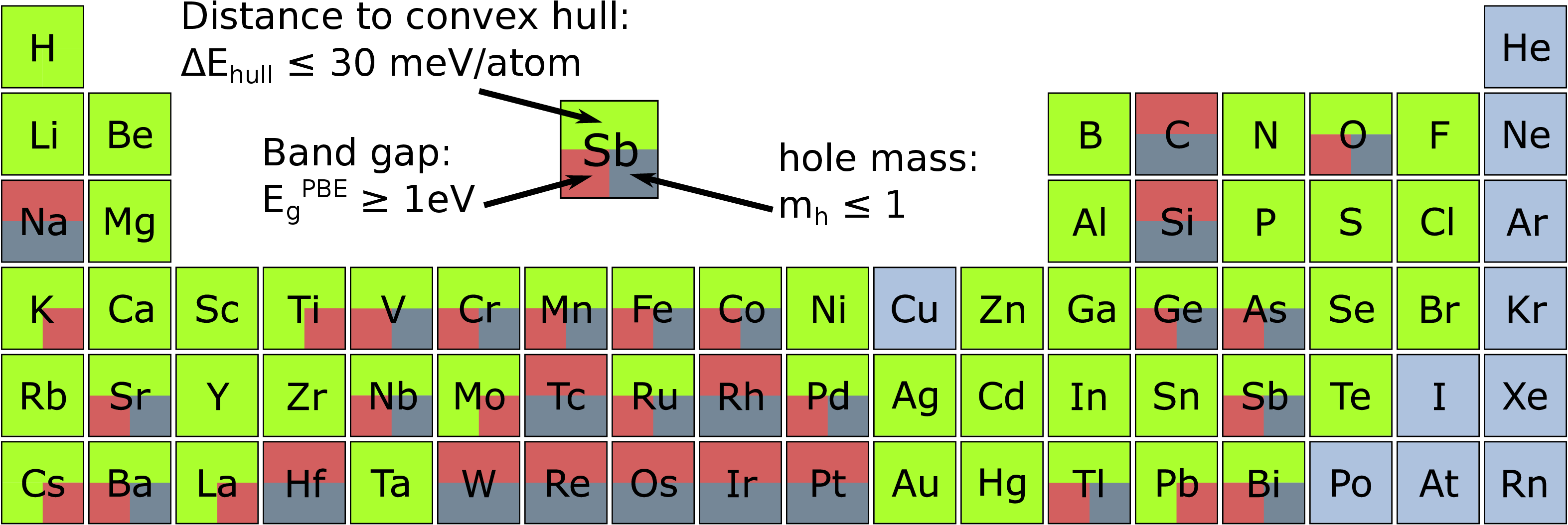}
\caption{Summary of the three filtering steps for all candidate alloying
  elements X. Top panel: stability results. Bottom left: transparency.
  Bottom right: hole mobility. A green tile indicates that at least one
  ternary compound containing that element has passed the corresponding
  filter step. A red tile indicates that no compound passed the step;
  the remaining filter results for such elements are greyed out.}
\label{Fig_PSE_CollectedData}
\end{figure*}

\section{High-throughput filtering}
\label{sec:results}

The goal of this search is to identify CuI-based ternary compounds that
combine the three key properties required for a p-type TCM:
thermodynamic stability, optical transparency in the visible range, and
low hole effective mass as a proxy for high hole mobility. Among the
candidate materials, we pay particular attention to p-type degenerate
semiconductors, defined here as compounds in which the Fermi level lies
slightly below the valence band maximum, yielding intrinsic hole
carriers without extrinsic doping, while a finite band gap still
separates the valence and conduction bands (see the Supporting
Information for the precise criterion). This combination is especially
attractive for device applications, as it removes the need for
controlled doping to achieve p-type conductivity. To identify such
materials systematically and without bias toward known structure types,
we apply a three-step sequential filter to the full set of structures
generated by the MHM: a thermodynamic stability filter, a transparency
filter based on the PBE band gap, and a hole mobility filter based on
the light hole effective mass.

The high-throughput workflow is shown in Fig.~\ref{Fig_Workflow}, and
each step is described in detail in the following subsections. The
results of the three steps are summarized in
Fig.~\ref{Fig_PSE_CollectedData}.

\subsection{Structure prediction and stability}

The periodic table visualization of Fig.~\ref{Fig_PSE_CollectedData}
highlights, for each candidate alloying element X, the lowest distance
to the convex hull found across all stoichiometries of the X--Cu--I phase
diagram. Structures with $\Delta E_\mathrm{hull}^\mathrm{PBE} <
30$~meV/atom passed the stability filter and were carried forward to the
electronic characterization step. This threshold is motivated by the
thermal energy at room temperature and is consistent with the criterion
applied in our earlier work~\cite{Schmidt_2022}. The full set of
individual PBE ternary phase diagrams for all considered elements X is
provided in Figs.~S3--S14 of the Supporting Information. Element-resolved
stability results evaluated with PBE and SCAN are shown in Fig.~S15 of
the Supporting Information.

\subsection{Electronic properties and final candidate set}

A total of 128 structures within 30~meV/atom of the convex hull were
characterized with respect to their electronic properties. The
element-resolved filtering results are shown in
Fig.~\ref{Fig_PSE_CollectedData}, where elements for which no stable
ternary was found are greyed out. More detailed numerical results are
provided in Fig.~S16 of the Supporting Information, where the color of
each tile reflects the best result observed among the remaining
structures for a given element. Here, ``best result'' is defined as
either the largest PBE band gap or, where present, p-type degenerate
semiconductor character combined with a large PBE band gap. The latter
is considered more favorable than a large band gap alone, as it implies
natural p-type carrier generation in the absence of doping. In total,
79 structures passed the transparency filter, and 20 of the 58 candidate
materials identified in this work are p-type degenerate semiconductors.

All structures with a PBE band gap below 1.0~eV and all metallic
structures were discarded at this stage. For structures passing the band
gap filter, light hole effective masses were computed. The results are
shown in Fig.~S16 as the lowest effective mass observed across all
Brillouin zone paths considered for each structure. The mobility filter
requires $m_\mathrm{lh} < 1.0\,m_0$ along at least one high-symmetry
path, which is more stringent than the spatially averaged threshold of
$1.5\,m_0$ used in Ref.~\cite{CEEM_2024}.

Applying the full three-step filter cascade yields 58 candidate p-type
TCMs. To our knowledge, only three of these have been previously reported
in the literature (excluding the sulfur- and selenium-based ternaries
published in our earlier work~\cite{Seifert_2024}):
CuGaI$_4$~\cite{MP_GaCuI4}, CuTe$_2$I~\cite{MP_Te2CuI}, and
Cu$_2$ZnI$_4$~\cite{Yamada_2022}, although in the case of
Cu$_2$ZnI$_4$ a slightly different crystal structure was reported. These
58 structures, along with their calculated band gaps and light-hole
effective masses, are summarized in Table~\ref{Table_58Structures}.
Additional details are provided in Tables~S1 and S2 of the Supporting
Information.

\begin{table*}[!htpb]
\centering
\caption{The 58 candidate p-type TCMs identified in this work, together with band gaps calculated with PBE and mHSE06, and the minimum light hole effective mass. For stoichiometries with multiple predicted phases, the space group number is given in parentheses. For the three compounds previously reported in the literature, the corresponding reference is provided. Additional computed properties are listed in Tables S1 and S2 of the Supporting Information.}
\vspace*{2mm}
\label{Table_58Structures}
\begin{tabular}{l c c c |@{\hspace{1cm}} l c c c}
Formula & $E_\mathrm{g}^\mathrm{PBE}$ (eV) &
$E_\mathrm{g}^\mathrm{mHSE}$ (eV) &
$m_\mathrm{lh}^\mathrm{PBE}$ ($m_0$) & Formula & $E_\mathrm{g}^\mathrm{PBE}$ (eV) &
$E_\mathrm{g}^\mathrm{mHSE}$ (eV) &
$m_\mathrm{lh}^\mathrm{PBE}$ ($m_0$) \\
\hline
H$_2$CuI$_2$ & 3.40 & 4.66 & 0.605 & Cu$_2$SeI & 1.30 & 2.60 & 0.355 \\
H$_2$CuI$_3$ & 2.31 & 4.09 & 0.497 & Cu$_3$SeI$_2$ & 1.20 & 2.20 & 0.263 \\
Li$_2$CuI$_3$ & 2.30 & 4.20 & 0.657 & CuTe$_2$I (Ref.~\cite{MP_Te2CuI}) & 0.91 & 2.48 & 0.361 \\
Li$_3$Cu$_2$I$_5$ & 2.36 & 4.21 & 0.563 & Cu$_2$FI$_2$ & 2.29 & 3.75 & 0.262 \\
LiCu$_2$I$_3$ & 1.73 & 3.59 & 0.452 & Cu$_2$Cl$_2$I & 2.49 & 3.92 & 0.587 \\
LiCuI$_2$ & 2.00 & 3.90 & 0.575 & Cu$_2$ClI$_2$ & 2.54 & 4.04 & 0.494 \\
Rb$_2$Cu$_3$I$_5$ & 1.91 & 3.55 & 0.928 & Cu$_2$Br$_2$I (spg 44) & 2.08 & 3.41 & 0.338 \\
BeCu$_2$I$_4$ & 1.86 & 3.54 & 0.378 & Cu$_2$Br$_2$I (spg 8) & 2.30 & 3.74 & 0.506 \\
MgCu$_2$I$_4$ & 2.10 & 3.71 & 0.519 & Cu$_2$BrI$_2$ (spg 44) & 2.41 & 3.88 & 0.490 \\
CaCu$_2$I$_4$ & 2.10 & 3.88 & 0.531 & Cu$_2$BrI$_2$ (spg 8) & 2.39 & 3.53 & 0.344 \\
Ca$_2$CuI$_5$ & 2.75 & 4.50 & 0.357 & Cu$_2$BrI & 1.63 & 3.44 & 0.478 \\
BCu$_3$I$_6$ & 1.95 & 3.60 & 0.459 & CuBrI & 2.43 & 3.26 & 0.310 \\
BCuI$_4$ & 2.38 & 4.53 & 0.631 & Sc$_2$CuI$_7$ & 1.51 & 2.94 & 0.610 \\
BCu$_2$I$_5$ & 2.07 & 3.57 & 0.525 & ScCu$_2$I$_5$ & 1.73 & 3.34 & 0.731 \\
Al$_2$CuI$_7$ & 2.49 & 4.16 & 0.593 & CuYI$_4$ & 1.35 & 2.83 & 0.389 \\
CuGa$_2$I$_7$ & 1.43 & 3.03 & 0.799 & CuZrI$_5$ & 1.74 & 2.97 & 0.357 \\
Cu$_2$GaI$_5$ (spg 1) & 0.99 & 2.40 & 0.459 & CuTaI$_6$ & 1.05 & 2.09 & 0.936 \\
Cu$_2$GaI$_5$ (spg 5) & 1.10 & 2.55 & 0.893 & NiCu$_2$I$_4$ & 1.59 & 2.14 & 0.552 \\
CuGaI$_4$ (spg 121) & 1.39 & 2.97 & 0.511 & NiCuI$_3$ & 1.83 & 2.06 & 0.344 \\
CuGaI$_4$ (spg 82, Ref.~\cite{MP_GaCuI4}) & 1.66 & 3.19 & 0.588 & CuAgI$_2$ & 1.00 & 2.77 & 0.233 \\
CuInI$_4$ (spg 121) & 1.15 & 1.62 & 0.611 & Cu$_2$AgI$_4$ & 2.03 & 3.37 & 0.380 \\
CuInI$_4$ (spg 82) & 1.36 & 2.78 & 0.789 & CuAg$_2$I$_3$ & 1.04 & 2.76 & 0.241 \\
CuSnI$_4$ & 2.38 & 3.20 & 0.622 & Cu$_2$AuI$_4$ & 1.40 & 2.69 & 0.350 \\
N$_2$CuI & 1.89 & 4.05 & 0.837 & Cu$_2$ZnI$_4$ (Ref.~\cite{Yamada_2022}) & 1.18 & 2.81 & 0.429 \\
PCuI$_4$ & 1.26 & 2.50 & 0.694 & CuZnI$_3$ & 1.49 & 3.20 & 0.439 \\
Cu$_3$SI$_2$ & 1.44 & 2.60 & 0.231 & Cu$_2$CdI$_4$ & 1.28 & 2.84 & 0.432 \\
Cu$_2$SI & 1.45 & 2.60 & 0.466 & CuCdI$_3$ & 1.24 & 2.89 & 0.433 \\
CuS$_2$I & 1.00 & 1.90 & 0.580 & CuCdI$_4$ & 1.95 & 2.99 & 0.353 \\
CuSe$_2$I$_5$ & 1.20 & 2.21 & 0.781 & CuHgI$_4$ & 1.10 & 2.00 & 0.415 \\
\end{tabular}
\end{table*}

\begin{figure}[htpb!]
\centering
\includegraphics[width=8.5cm]{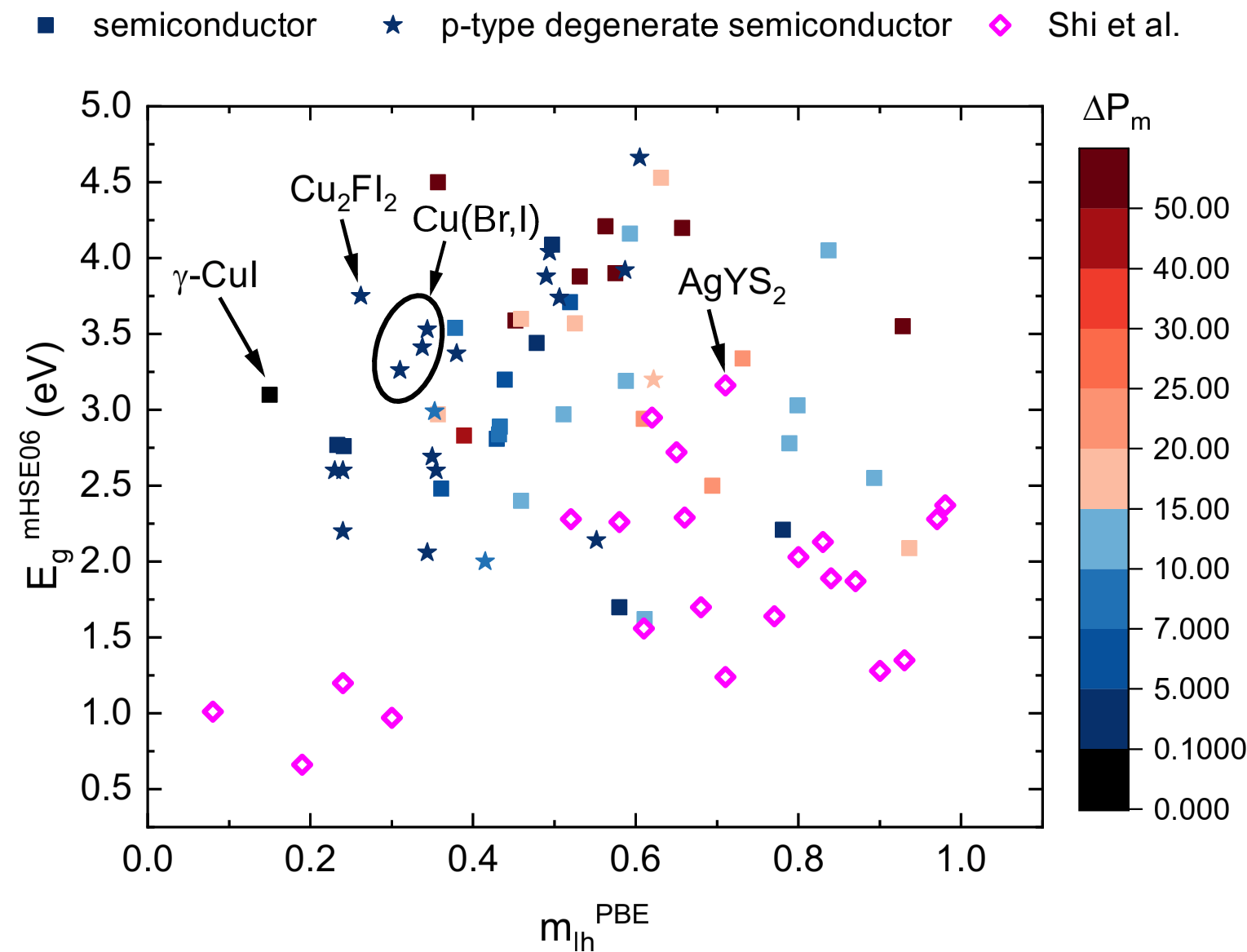}
\caption{Band gap calculated with mHSE06 versus minimum light hole effective mass for
  the 58 candidate p-type TCMs. Symbol color encodes the distance to Cu or I on the P$_\mathrm{m}$ scale introduced in Ref.~\cite{Glawe_2016}, which was derived by a statistical analysis of the likelihood of a chemical element being replaced by another; symbol shape distinguishes conventional semiconductors
  (circles) from p-type degenerate semiconductors (triangles). For comparison, we added data from Shi et al.~\cite{Shi_2017}, another high-throughput study of p-type TCMs. Please note that the band gaps given by Shi et al.~\cite{Shi_2017} are calculated with HSE06 instead of mHSE06. Some of the materials proposed therein have larger hole effective masses and are therefore not displayed here. The three structures noted as Cu(Br,I) alloys are CuBrI, Cu$_2$Br$_2$I (space group 44), and Cu$_2$BrI$_2$ (space group 8).}
\label{Fig_GapQSCSim}
\end{figure}

The distribution of band gaps and effective masses is shown in
Fig.~\ref{Fig_GapQSCSim}. mHSE06 band gaps range from approximately
1.6 to 4.7~eV across the 58 candidates, with p-type degenerate
semiconductors (triangles) concentrated at lower effective masses on
average. This reflects the tendency of Cu-deficient or
mixed-halide stoichiometries, which favor natural hole generation, to
also support dispersive valence bands. A transparency concern specific to p-type degenerate semiconductors is free-carrier (Drude-like) absorption and, more generally, optical transitions involving the partially empty valence bands, analogous to the ``second gap'' transitions discussed for n-type transparent conducting oxides in Ref.~\cite{Ha_2016}. These contributions were not computed here; this question is revisited in the
context of the design rules in Sec.~\ref{sec:rule5}.
Compared to the structures proposed by Shi et al.~\cite{Shi_2017}, both the transparency and the hole mobility of our CuI-based alloys are superior: the four structures from that study with a remarkably low hole effective mass of around $0.2\,m_0$ all have band gaps below 1.5~eV, limiting their transparency. Finally, regarding the alloys predicted here, we color-coded the distance of the third element X to either Cu or I using the data-mined P$_\mathrm{m}$ scale introduced in Ref.~\cite{Glawe_2016}. This scale indicates the likelihood of one chemical element replacing another. Outliers are Li and Rb, which are far away from Cu on this scale. This can be related to the ionic nature of our systems, as well as to the fact that we have Cu$^\mathrm{+}$ instead of Cu$^\mathrm{2+}$ in the case of CuI. This is discussed in more detail in Sec.~\ref{sec:rule5}. The best material with a large distance of the third element to Cu and I is Ca$_2$CuI$_5$ with a band gap of 4.5~eV and a hole effective mass of $0.357\,m_0$. 

While the full structural dataset is hosted in the {\sc Alexandria}
database~\cite{Schmidt_2022, Schmidt_2023, Schmidt_2024, Cavignac_2026},
the specific property values discussed in this work are provided in the
Supporting Information (Tables~S1 and S2). These 58 candidate materials
serve as the foundation for the design rule analysis presented in the
following section.

\section{Revisiting the CMVB design rules}
\label{sec:design}

The 58 candidate p-type TCMs identified in the preceding section provide
a structurally diverse, non-oxide dataset with which to systematically
evaluate the design rules for p-type TCMs originally proposed in the
framework of chemical modulation of the valence band
(CMVB)~\cite{Kawazoe_1997, Kawazoe_2000}. The CMVB rules were
formulated for Cu-based delafossite oxides with linearly coordinated
Cu$^+$ (coordination number CN~=~2) and O~2$p$ as the relevant anion
states. The present dataset differs from this original context in two
important respects: the anion is iodine (I~5$p$) rather than oxygen
(O~2$p$), and the dominant coordination environment of Cu is tetrahedral
(CN~=~4) rather than linear. It is therefore non-trivial whether, and
to what extent, the CMVB rules transfer to this broader chemical space.

Before proceeding, three caveats regarding the dataset should be kept in
mind. First, all structures contain both Cu and I by construction, which
introduces a systematic bias toward certain bonding environments.
Second, all 58 structures share the three properties used as filter
criteria (stability, transparency, and low hole effective mass), so any
feature common to the set is at most a necessary condition for p-type
TCM character, not a sufficient one. Third, the dataset lacks contrast:
the absence of failed candidates makes it impossible to definitively
distinguish causal factors from coincidental correlations. With these
limitations in mind, we address each CMVB design rule in turn in the
following subsections. A summary of the verdicts is provided in
Table~\ref{Table_CompCMVB}.

\subsection{Rule 1: Ionic bonding and the closed-shell
  Cu\textsuperscript{+} cation}
\label{sec:rule1}

\begin{figure}[htpb!]
\centering
\includegraphics[width=\columnwidth]{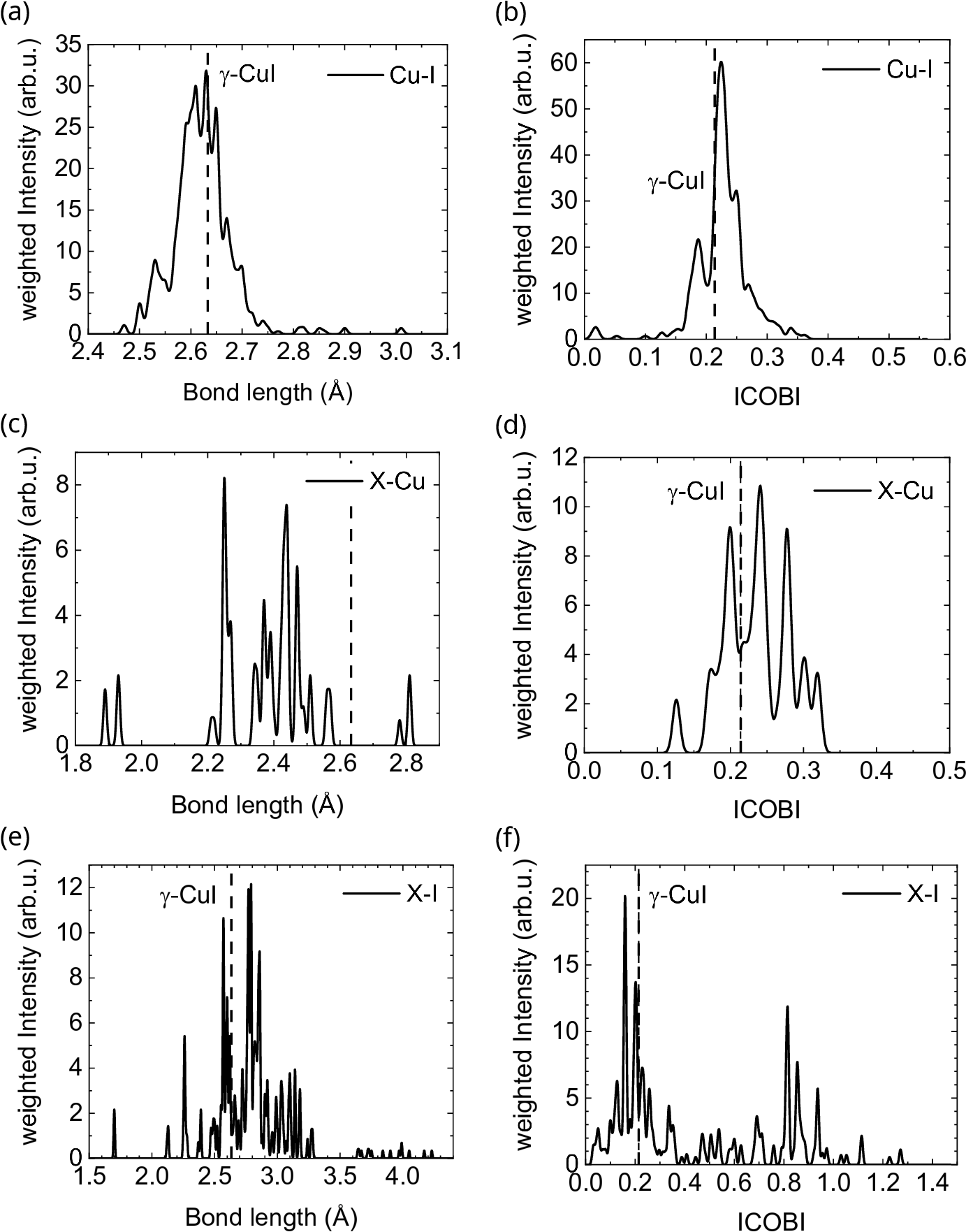}
\caption{Distribution of bond lengths (a, c, e) and ICOBI bonding
  indices (b, d, f) for Cu--I (a, b), X--Cu (c, d), and X--I (e, f)
  bonds in the dataset of 58 candidate p-type TCMs. Bond lengths and
  ICOBI values were broadened by 0.005~\AA\ and 0.005, respectively, to
  form a continuous spectrum. ICOBI values were computed with PBE using
  LOBSTER~\cite{Lobster}. }
\label{Fig_BLICOBI1}
\end{figure}

The first CMVB requirement is that the active cation be a closed-shell
$d^{10}$ species (Cu$^+$ in the original formulation) embedded in an
ionic crystal where the valence states are well separated from the
conduction bands of the host. Ionic bonding at the Cu site is therefore
a prerequisite for the CMVB mechanism to operate.

The bonding type in the dataset was characterized using the ICOBI
index~\cite{Ertural_2021}, with results shown in Fig.~\ref{Fig_BLICOBI1}
for Cu--I, X--Cu, and X--I bonds; X--X and I--I bonds are shown in
Fig.~S21 of the Supporting Information. The Cu--I bond lengths and ICOBI
values cluster closely around the values of $\gamma$-CuI, confirming
that the local ionic bonding environment of CuI is largely preserved
across the ternary compounds. The low ICOBI values for Cu--I bonds are
consistent with predominantly ionic character, in agreement with the
literature on CuI~\cite{Phillips_1970}.

For X--Cu bonds, ICOBI values remain similarly low on average,
indicating that X generally adopts a role consistent with the ionic
lattice. For X--I bonds, a broader distribution is observed, with some
structures showing higher ICOBI values indicative of increased
covalency. This is rationalized by vacancy formation in compounds such
as CuGaI$_4$, where the higher oxidation state of Ga relative to Cu
reduces the number of bonds, forcing the remaining ones to carry more
bonding density. X--X and I--I bonds are rare throughout the dataset;
the largest X--X ICOBI value originates from the triple bond in
N$_2$CuI. No Cu--Cu bonds are present in any of the 58 candidates.

The overall picture confirms that the ionic bonding environment of CuI
is maintained across the ternary compounds: the third element X
substitutes into the ionic lattice either on the cation or anion
sublattice, with Cu--I, X--Cu, and X--I bonds dominating.

Regarding the specific requirement of Cu$^+$ ($d^{10}$) character: this
is implicitly enforced by the ionic bonding environment and the dominant
+1 oxidation state of Cu in CuI-derived structures. For the majority of
the dataset, the stoichiometric formula together with the computed
electronic structure confirms Cu$^+$ rather than Cu$^{2+}$ character.
For the Ni-containing compounds (NiCu$_2$I$_4$ and NiCuI$_3$), charge balance implies Ni$^{2+}$ with an open $3d^8$ shell and Cu in the +1 oxidation state. This is consistent with the ferromagnetic ground state obtained in our spin-polarized calculations and observed in synthesized zincblende-based Ni$_x$Cu$_{1-x}$I alloys~\cite{Dethloff_2024}. The Ni $d$ states mix with the Cu $d$ manifold near the VBM. For Ag-based compounds, the $d^{10}$ character of Ag$^+$
is analogous to that of Cu$^+$, and $p$-$d$ hybridization is similarly
operative, as evidenced by the dispersive valence bands reported for
Ag$_x$Cu$_{1-x}$I alloys~\cite{Krueger_2023_AgCuI}.

We conclude that Rule~1 is confirmed for the large majority of the
dataset. The ionic bonding environment of CuI is preserved in the
ternary compounds, consistent with the $d^{10}$ Cu$^+$ requirement.

\subsection{Rule 2: Coordination environment of Cu}
\label{sec:rule2}

\begin{figure}[!htbp]
\begin{center}
\includegraphics[width=0.9\columnwidth]{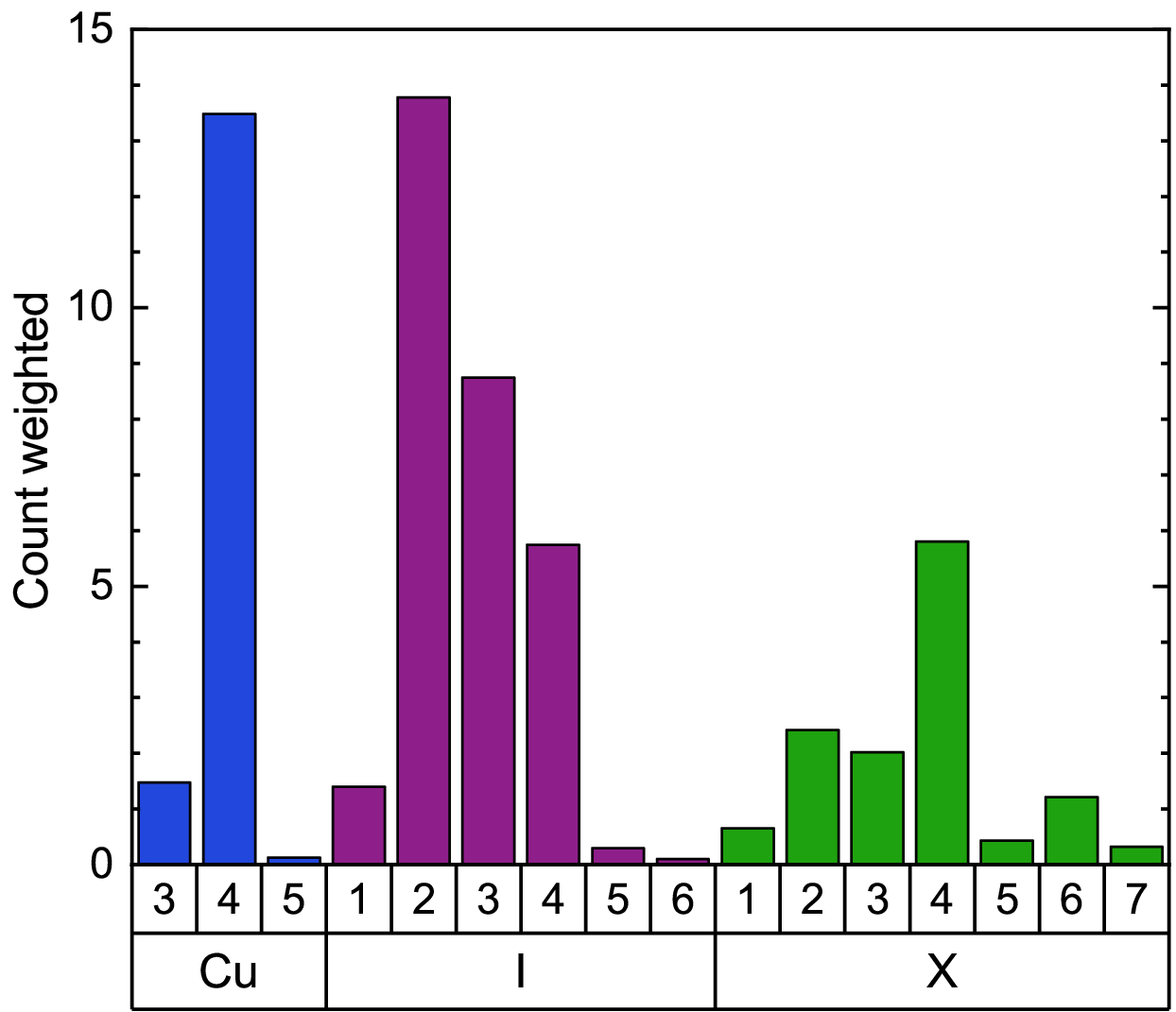}
\caption{Occurrence of coordination numbers in the 58 candidate p-type
  TCMs. Results are weighted such that every material contributes unit
  total intensity to the distribution, properly accounting for
  structures with different numbers of atoms in the unit cell.}
\label{Fig_CoordinationNumber}
\end{center}
\end{figure}

In the original CMVB framework, Cu is linearly coordinated to two
oxygen atoms in an O--Cu--O dumbbell motif (CN~=~2), as in the
delafossite structure. It has been argued that this specific
coordination geometry promotes the orbital overlap between Cu~$3d$ and
O~$2p$ states that gives rise to the dispersive
VBM~\cite{Williamson_2017}. An important question is therefore whether
CN~=~2 is a necessary condition, or whether other coordination
environments can support p-type TCM character.
Ref.~\cite{Williamson_2017} reported coordination numbers greater than
2 for the optimal structures in X--Cu--P alloys (X = Mg, Ca, Ba, Sr),
indicating that the ideal coordination number depends on the specific
material class.

The coordination number distributions for Cu, I, and X in the dataset
are shown in Fig.~\ref{Fig_CoordinationNumber}, weighted such that each
material contributes unit total intensity regardless of the number of
atoms in its unit cell. The most striking observation is that Cu almost
universally adopts CN~=~4 (tetrahedral coordination), as in
$\gamma$-CuI itself. Linear coordination (CN~=~2), which characterizes
the delafossite structure, is not observed as the dominant motif in any
compound in the dataset.

For iodine, the distribution peaks at smaller coordination numbers,
reflecting two effects: the formation of cation vacancies in some
structures (e.g., CuGaI$_4$), and the prevalence of layered structures
in which I atoms located at the outer surface of a layer have fewer
neighbors than in a three-dimensionally bonded crystal. For X, the
coordination number distribution is broad, consistent with the chemical
diversity of the third element. The space group distribution of the 58
candidates, provided in Table~S4 of the Supporting Information, confirms
that the MHM explores well beyond the zincblende-derived structure types
(space groups 44 and 82), with a majority of structures in low-symmetry
space groups (space groups 1 and 8), reflecting the structural diversity
of the predicted compounds.

The dominance of CN~=~4 for Cu across the dataset shows that tetrahedral coordination is compatible with p-type TCM character in CuI-based ternary compounds; linear coordination as in the delafossites (CN~=~2) is therefore not a necessary condition. Since all candidates share this feature, the present dataset cannot establish whether CN~=~4 is by itself sufficient. A further pattern evident in
Fig.~\ref{Fig_CoordinationNumber} is that a non-negligible fraction of
Cu sites, and more prominently I sites, show coordination numbers below
4. This tendency toward undercoordination reflects the ease of Cu
vacancy formation in the iodide lattice. This is a consequence of the
monovalent I$^-$ anion imposing weaker electrostatic constraints on the
removal of a Cu$^+$ cation than the divalent O$^{2-}$ anion does in
oxide-based systems. Cu vacancy formation is already the dominant
intrinsic defect mechanism in $\gamma$-CuI, responsible for its natural
p-type carrier
concentration~\cite{Wang_2011, Huang_2012, Grauzinyte_2019}. The same
tendency reappears structurally in ternary compounds such as
CuGaI$_4$, where the substitution of Ga$^{3+}$ for Cu$^+$ is
crystallographically equivalent to a local Cu-deficient environment.

As a result, Rule~2 is generalized in two respects. First, linear coordination (CN~=~2) is not required: tetrahedral Cu coordination (CN~=~4) is compatible with p-type TCM character. Second, a structural tendency toward Cu
undercoordination, manifested as Cu vacancy formation and Cu-deficient
stoichiometries, appears to be a favorable feature rather than a
deficiency, as it is directly connected to natural p-type carrier
generation. This point is developed further in Rule~6
(Sec.~\ref{sec:rule6}).

\subsection{Rule 3: $p$-$d$ hybridization at the VBM}
\label{sec:rule3}

\begin{figure*}[htpb!]
\centering
\includegraphics[width=16.6cm]{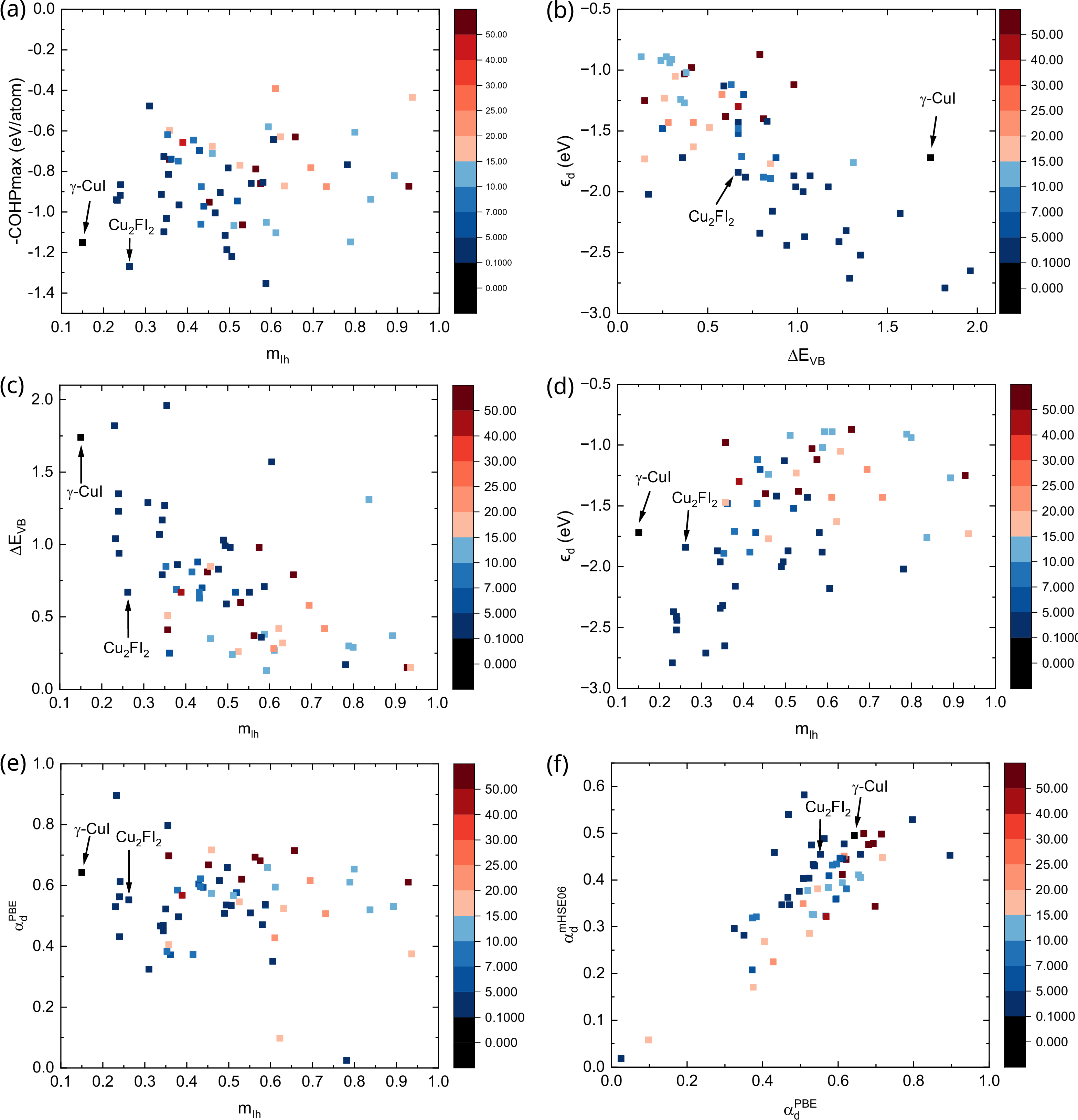}
\caption{Valence-band descriptors for the 58 candidate p-type TCMs.
  (a) Maximum of $-\mathrm{COHP}$ close to the VBM versus the light
  hole effective mass, following the antibonding-state metric of
  Ref.~\cite{Das_2023}. (b) Center of the Cu $d$ states relative to
  the VBM, $\epsilon_d$, versus the topmost valence band width
  $\Delta E_\mathrm{VB}$. (c) $\epsilon_d$ and (d) $\Delta E_\mathrm{VB}$
  versus the light hole effective mass. (e) $d$ state contribution at
  the VBM, $\alpha_d$, versus the light hole effective mass computed
  with PBE; marginal distributions are shown as kernel density estimates
  with a broadening of 0.005. (f) Comparison of $\alpha_d$ computed
  with PBE and mHSE06. All quantities were obtained with PBE except
  where mHSE06 is indicated; COHP values were computed with
  LOBSTER~\cite{Lobster}. The color map encodes the distance of the third element X to either Cu or I, as in Fig.~\ref{Fig_GapQSCSim}.}
\label{Fig_COHP}
\end{figure*}

The central mechanism of CMVB is the hybridization of the Cu $3d$
states with the anion $p$ states at the VBM. This hybridization
disperses the VBM, lowers the hole effective mass, and simultaneously
pushes antibonding states to the top of the valence band. Two
complementary quantities are used here to characterize this: the
$d$ state contribution at the VBM ($\alpha_d$,
Fig.~\ref{Fig_COHP}(e)), and the COHP antibonding character near the
VBM (Fig.~\ref{Fig_COHP}(a)).

The contribution of $d$ states at the VBM, $\alpha_d$, is shown versus
the light hole effective mass in Fig.~\ref{Fig_COHP}(e). Almost all
structures show a substantial $d$ state contribution, with negligible
$s$-state weight at the VBM, confirming that $p$-$d$ hybridization is
universally present in the candidate set. The values of $\alpha_d$ are
broadly distributed within the interval $(0.3, 0.7)$, indicating that
the hybridization is significant but not dominant. The only compound
exhibiting a notable $s$-state contribution at the VBM is CuSnI$_4$,
for which the Sn $5s$ states participate in the uppermost valence band.
No direct correlation between $\alpha_d$ and the light hole effective
mass is observed, consistent with $p$-$d$ hybridization being a
necessary but not sufficient condition for low hole effective mass.

The absence of correlation is partly attributable to the dataset bias:
since all structures contain both Cu and I, some degree of $p$-$d$
hybridization is expected even in wide-gap insulating phases of these
elements, making it difficult to distinguish the contribution of
hybridization from other structural factors. The data are therefore
consistent with $p$-$d$ hybridization being a necessary ingredient, but
the dataset alone cannot establish sufficiency.

Because PBE is known to underestimate the localization of Cu $d$ states,
we also determined $\alpha_d$ with mHSE06. As shown in
Fig.~\ref{Fig_COHP}(f), the mHSE06 values are systematically shifted
toward lower $d$ state contributions relative to PBE, but the two sets
of values are strongly correlated, confirming that the qualitative
picture is insensitive to the choice of functional.

The trace of $p$-$d$ hybridization is independently visible in the COHP
analysis. All 58 structures exhibit antibonding states close to the VBM
(Fig.~\ref{Fig_COHP}(a)), the expected fingerprint of $p$-$d$
hybridization pushing filled antibonding combinations to the top of the
valence band. The magnitude of the negative COHP at the VBM shows
considerable scatter versus the light hole effective mass and does not
reveal a simple linear correlation. COHP-based analysis of antibonding
states at the VBM has been used as a descriptor for thermoelectric
materials~\cite{Yuan_2023, Ubaid_2024} and was recently applied to the
specific case of CuBiI$_4$~\cite{Das_2023}. To our knowledge, its
systematic evaluation across a p-type TCM candidate set is new.

We conclude that Rule~3 is confirmed and extended. The $p$-$d$
hybridization is universally present in the dataset, with I~$5p$ rather
than O~$2p$ as the relevant anion $p$ states. Its presence is evidenced
both directly through $\alpha_d$ and indirectly through antibonding
states in the COHP. However, the value of $\alpha_d$ alone does not
determine the hole effective mass; additional structural and chemical
factors are required.

\subsection{Rule 4: Valence band dispersion and the role of the
  $d$ state center}
\label{sec:rule4}

Beyond the question of whether $p$-$d$ hybridization is present,
the CMVB framework implicitly requires that the Cu $d$ states be
energetically close to the anion $p$ states. If the $d$ states lie too
deep in energy, hybridization is suppressed; if they lie too high, the
compound may become metallic. This suggests that the energetic position
of the $d$ states provides a quantitative descriptor for valence band
dispersion, and hence for the hole effective mass, beyond the qualitative
presence of hybridization.

We test this using the center of the Cu $d$ states relative to the VBM,
$\epsilon_d$, and the width of the topmost valence band across the full
Brillouin zone, $\Delta E_\mathrm{VB}$. Both quantities are plotted
against each other and against the light hole effective mass in
Figs.~\ref{Fig_COHP}(b)--(d). The $d$-band center model has been
established as a predictor of surface reactivity and chemisorption in
transition metals~\cite{Hammer_1996, Ruban_1997, Christoffersen_2001};
here we explore its transferability to valence band dispersion in p-type
semiconductors.

The key finding is a strong correlation between $\epsilon_d$ and
$\Delta E_\mathrm{VB}$ (Pearson $r = -0.75$; see Table~S3 of the
Supporting Information for the full correlation matrix). Materials in
which the Cu $d$ states deeper from the VBM tend to have a broader
topmost valence band (meaning they are spanning a large energy distance over the Brillouin zone), consistent with stronger hybridization and greater
band dispersion across the Brillouin zone. The correlations of both
$\epsilon_d$ and $\Delta E_\mathrm{VB}$ with the light hole effective
mass are moderate ($r = 0.53$ and $-r = -0.54$, respectively). The weaker
correlation with the effective mass, compared to the strong
$\epsilon_d$--$\Delta E_\mathrm{VB}$ correlation, reflects the fact
that the effective mass probes only a small region of $k$-space around
the VBM, whereas $\Delta E_\mathrm{VB}$ captures the overall valence
band dispersion. The $d$ state center is therefore a better predictor of
the global valence band width than of the local curvature at the VBM.

This constitutes a quantitative refinement of the CMVB prescription:
the energetic position of the Cu $d$ states matters, not just their
presence at the VBM. Compounds in which $\epsilon_d$ lies closer to the
VBM are predicted to show broader valence bands and, on average, lower
hole effective masses, making $\epsilon_d$ a useful pre-screening
descriptor for p-type TCM candidates. As shown in Fig.~S22(b) of the Supporting Information, $\epsilon_d$ is moderately correlated with the
Pauling electronegativity of X ($r = -0.56$): more electronegative
third elements tend to pull the Cu $d$ states to deeper energies,
reducing hybridization. This observation provides a chemical handle for
tuning valence band dispersion through the choice of X. Individual
$\epsilon_d$ values are tabulated in Tables~S1 and S2.

We conclude that Rule~4 is confirmed and quantified. The energetic
position of the Cu $d$ states ($\epsilon_d$) is a strong predictor of
the overall valence band width ($r = 0.75$) and a moderate predictor of
the light hole effective mass ($r = 0.53$), and the Pauling
electronegativity of X provides a chemically intuitive handle for
tuning it.

\subsection{Rule 5: Wide-gap second cation and transparency}
\label{sec:rule5}

The CMVB framework requires the second cation (element X in our
notation) to have a large band gap in its own compounds, ensuring that
the X sublattice does not introduce low-energy states that would reduce
the optical transparency of the ternary compound. To test this, we
examined the band gaps of the relevant binary X--I and Cu--X phases
(see Tables~S5 and S6 of the Supporting Information). For X--I binary
phases, band gaps are uniformly large across the set of candidate X
elements. For Cu--X binary phases, the situation is more nuanced:
Cu--chalcogenide binaries have relatively small band gaps, though the
presence of iodine in the ternary appears to restore a large optical
gap in most cases. In the present high-throughput workflow, this rule is
in part implicitly enforced by the band gap filter
($E_\mathrm{g}^\mathrm{PBE} > 1.0$~eV).

The mHSE06 band gaps for the 58 candidates span the range 1.6--4.7~eV
(Table~\ref{Table_58Structures}; the full gap distribution is shown in
Fig.~S22(a) of the Supporting Information). Thirty compounds have mHSE06 gaps of at least 3.1~eV, comparable to $\gamma$-CuI, and are expected to be transparent across the entire visible range. The remaining 28 have gaps between 1.6 and 3.1~eV, corresponding to absorption onsets between 775~nm and 400~nm, and therefore absorb part of the visible spectrum; for these compounds, adequate transparency can only be expected in thin films.

Among the 58 candidates, 20 are p-type degenerate semiconductors, with
third elements drawn from H, Sn, S, Se, F, Cl, Br, Ni, Ag, Au, Cd, and
Hg (see Table~\ref{Table_58Structures}).

Most of these elements are chemically close to either Cu or I: since the compounds are predominantly ionic, alloying occurs either on the anion or the cation sublattice. The likelihood of one element replacing another was statistically investigated by some of us in Ref.~\cite{Glawe_2016}, yielding the P$_\mathrm{m}$ scale, which reorders the elements of the periodic table accordingly. On this scale, elements close to I include the other halogens and the chalcogens as well as H, which matches intuition. Considering Cu as a transition metal, the picture is more complex. In the direct neighborhood one finds Ag, Au, and Ni, while Zn, Cd, and Hg are up to eight positions away and therefore also close to Cu. The outlier is Sn. As shown in Fig.~\ref{Fig_GapQSCSim},
compounds in which X is chemically similar to either Cu or I show lower
hole effective masses on average.

The identification of 20 p-type degenerate semiconductors extends the
picture: these materials combine natural hole carriers with a finite
gap, making them particularly attractive as p-type TCMs. Transparency of this subclass requires that neither free-carrier (Drude) absorption nor transitions involving the partially empty valence bands extend into the visible range. For n-type transparent conducting oxides, the analogous transitions are governed by the so-called second gap~\cite{Ha_2016}. For the compounds identified here, whose formal hole concentration can reach one hole per formula unit (see Sec.~\ref{sec:rule6}), a quantitative assessment of these contributions requires explicit calculations of the optical response and remains an open question.

We conclude that Rule~5 is broadly satisfied for the identified
candidates, primarily because the band gap filter applied during the
workflow ensures that transparent compounds are selected. The X--I
binary gaps support this conclusion; the Cu--X binaries, particularly
for chalcogenide X, represent a partial exception where the iodine
content of the ternary appears to compensate for a smaller binary gap.

\subsection{Rule 6: P-type dopability and natural carrier generation}
\label{sec:rule6}

The original CMVB framework was motivated in part by the difficulty of
achieving p-type doping in conventional oxides, where the localized
O~$2p$ VBM leads to deep acceptor levels and large hole effective
masses. A dispersive VBM obtained through CMVB lowers the acceptor
formation energy and enables shallower dopant levels, facilitating
p-type carrier generation.

The connection to Rule~2 is direct: the same tendency toward Cu
undercoordination that is structurally expressed as a coordination
number below 4 for some Cu sites is electronically expressed as hole
generation. In the ionic picture, removing a Cu$^+$ from the lattice
leaves behind an uncompensated negative charge on the surrounding
I$^-$ sublattice, which is equivalent to introducing a hole in the
valence band. The tendency of the iodide lattice to accommodate Cu
undercoordination, due to the relatively low electrostatic penalty of
removing a monovalent cation from a monovalent-anion host, thus
constitutes a structural design principle specific to halide-based
p-type TCMs and has no direct counterpart in the oxide-based CMVB
framework.

A compound whose stoichiometry places it in a Cu-deficient charge
balance will naturally generate holes without any externally introduced
doping. This is the defining property of the 20 p-type degenerate
semiconductors identified in this work. A concrete example is provided
by CuCdI$_4$: with Cd$^{2+}$, Cu$^+$, and I$^-$, the formal charge
count gives $+2 + 1 - 4 = -1$, which cannot be satisfied with integer
oxidation states and instead generates one hole per formula unit. An
analogous argument applies to Cu$_2$FI$_2$, where the highly
electronegative F$^-$ imposes a charge balance of
$2(+1) + (-1) + 2(-1) = -1$. The mechanism responsible for
natural p-type behavior can therefore be read directly from the
stoichiometric formula using standard oxidation state arguments. The
individual band structures of the 20 p-type degenerate semiconductors
are displayed in Figs.~S17--S20 of the Supporting Information.

For the parent compound $\gamma$-CuI, a systematic computational study
of 64 substitutional impurities has established that chalcogen elements
(S, Se) on the iodine sublattice are the most effective p-type dopants,
and that eight elements are viable n-type
dopants~\cite{Grauzinyte_2019}. For the CuI ternaries with S and Se
studied in our earlier work~\cite{Seifert_2024}, carrier concentrations
were shown to be tunable via Cu/I stoichiometry, and Cu$_2$SI,
Cu$_2$SeI, and Cu$_3$SI$_2$ were identified as p-type degenerate
semiconductors. The 20 p-type degenerate semiconductors identified in
the present work go one step further: they achieve natural p-type
carrier generation without any extrinsic doping, by virtue of their
stoichiometry placing the Fermi level below the VBM. This represents
the most favorable scenario from a device perspective, as it removes
the need for controlled doping.

We conclude that Rule~6 is supported by prior work for the parent CuI
system~\cite{Grauzinyte_2019, Seifert_2024} and extended by the present
dataset. The 20 p-type degenerate semiconductors represent a
particularly attractive subclass for which natural hole generation is
achieved without extrinsic doping, predictable from stoichiometric
charge balance arguments.

\subsection{Rule 7: Cu--Cu distance and $d$--$d$ transitions}
\label{sec:rule7}

In the original CMVB analysis of delafossite oxides, the Cu--Cu
distance was identified as a relevant parameter: if Cu atoms are too
close, $d$--$d$ transitions between neighboring Cu sites can introduce
absorption in the visible range and compromise
transparency~\cite{Jansen_1987}. A sufficient Cu--Cu distance is
therefore implicitly required for optical transparency.

In the present dataset, this rule is generally satisfied: no strong
Cu--Cu bonds are present in any of the 58 candidate structures. Eight
compounds (Rb$_2$Cu$_3$I$_5$, MgCu$_2$I$_4$, CaCu$_2$I$_4$,
BCu$_3$I$_6$, BCu$_2$I$_5$, Cu$_2$BrI, Cu$_2$ZnI$_4$,
Cu$_2$CdI$_4$) exhibit short Cu--Cu contacts arising from structural
competition between disparate binary phases; their minimal distances and
ICOBI values are listed in Table~S7, and the full distance distribution
is shown in Fig.~S23 of the Supporting Information. In all cases the
ICOBI values for Cu--Cu pairs are significantly lower than those of the
Cu--I bonds, confirming that these are near-neighbor contacts rather
than formal bonding partners. For iodide-based systems, optical
transitions near the gap edge are predominantly of Cu~$3d$--I~$5p$
character rather than Cu~$3d$--Cu~$3d$ character, making this rule less
critical in the halide context than in the original oxide
framework~\cite{Jansen_1987}.

We conclude that Rule~7 is satisfied throughout the dataset.

\subsection{Summary of design rules}
\label{sec:rulesum}

Table~\ref{Table_CompCMVB} summarizes the verdict for each CMVB design
rule as assessed against the present dataset. Per-material verdicts for
all 58 candidates are provided in Table~S8 of the Supporting
Information.

\begin{table*}[htb]
\centering
\caption{Assessment of the CMVB design rules for p-type TCMs against
  the present dataset of 58 CuI-based ternary compounds. ``Confirmed''
  indicates direct correspondence with the original rule;
  ``generalized'' indicates that the rule is satisfied in a broader form than originally stated.}
\vspace*{2mm}
\label{Table_CompCMVB}
\begin{tabular}{p{5.0cm} | p{5.0cm} | p{4.5cm} | p{3.0cm}}
Original CMVB rule & Test in this work & Result & Key reference \\
\hline
Closed $d^{10}$ shell of active cation &
  ICOBI bonding index; oxidation-state analysis (Rule~1) &
  confirmed &
  Fig.~\ref{Fig_BLICOBI1}, Fig.~S21 \\
\hline
Cu linearly coordinated (CN~=~2) &
  Coordination number distribution (Rule~2) &
  generalized (CN~=~2 not required; CN~=~4 compatible) &
  Fig.~\ref{Fig_CoordinationNumber} \\
\hline
Hybridization of Cu~$3d$ with anion~$p$ at VBM &
  $\alpha_d$, COHP (Rule~3) &
  confirmed (I~$5p$ replaces O~$2p$) &
  Fig.~\ref{Fig_COHP}(a, e) \\
\hline
Energetic proximity of Cu~$d$ and anion~$p$ &
  $\epsilon_d$ vs.\ $\Delta E_\mathrm{VB}$ (Rule~4) &
  generalized (quantitative descriptor) &
  Fig.~\ref{Fig_COHP}(b--d), Fig.~S22(b) \\
\hline
Large band gap of second cation &
  Binary phase gaps of X (Rule~5) &
  confirmed (partly enforced by the band-gap filter; caveats for Cu--X chalcogenides) & Fig.~S22(a), Tables~S5, S6 \\
\hline
P-type dopability via dispersive VBM &
  Charge-balance argument; degenerate SCs (Rule~6) &
  generalized (stoichiometric hole generation) & Figs.~S17--S20 \\
\hline
Sufficient Cu--Cu distance &
  ICOBI for Cu--Cu contacts (Rule~7) &
  confirmed &
  Fig.~S23, Table~S7 \\
\hline
\end{tabular}
\end{table*}

Four generalizations of the original CMVB framework emerge from the
analysis. First, linear coordination as in the delafossites (CN~=~2) is not required: tetrahedral Cu coordination (CN~=~4) is compatible with p-type TCM character. Second, the relevant anion $p$ states are
I~$5p$ rather than O~$2p$, demonstrating that the CMVB mechanism is
not restricted to oxide chemistries and extends naturally to
halide-based systems. Third, the energetic position of the Cu $d$
states ($\epsilon_d$) provides a quantitative descriptor for valence
band dispersion that refines the qualitative CMVB prescription into a
predictive tool. Materials with $\epsilon_d$ more far away from the VBM tend to
exhibit broader valence bands and lower hole effective masses, and the
Pauling electronegativity of X offers a handle for tuning this
quantity. Fourth, the tendency toward Cu undercoordination,
structurally expressed as Cu vacancy formation and Cu-deficient
stoichiometries, is an intrinsic feature of halide-based systems rather
than a defect, arising from the low electrostatic penalty for removing
a monovalent Cu$^+$ cation from a monovalent-anion host. This directly
links the coordination environment of Cu to natural p-type carrier
generation, unifying Rules~2 and~6 into a single design principle with
no direct counterpart in the oxide-based CMVB framework.

\section{Conclusion}
\label{sec:conclusion}

We have performed a high-throughput computational search for CuI-based
ternary compounds as candidate p-type transparent conducting materials.
Using the minima hopping method combined with density functional theory,
we explored the ternary phase diagrams of X--Cu--I systems for a broad
selection of elements X from the periodic table. Applying sequential
filters for thermodynamic stability, optical transparency, and light
hole effective mass, we identified 58 candidate p-type TCMs, of which
only three are known in the literature, excluding the sulfur- and
selenium-based ternaries reported in our earlier
work~\cite{Seifert_2024}. The full dataset is made available through
the {\sc Alexandria}
database~\cite{Schmidt_2022, Schmidt_2023, Schmidt_2024, Cavignac_2026}.
Among the 58 candidates, 20 are p-type degenerate semiconductors,
combining a finite band gap with intrinsic hole carriers in the absence
of extrinsic doping.

The central contribution of this work goes beyond the discovery of new
candidate materials. We used the dataset as a testing ground to
systematically evaluate the design rules of the chemical modulation of
the valence band (CMVB) framework~\cite{Kawazoe_1997, Kawazoe_2000},
originally formulated for Cu-based delafossite oxides with linear Cu
coordination and O~$2p$ anion states. Four generalizations of the
original framework emerge from the analysis.

First, linear coordination (CN~=~2) as in the delafossite structure is not required: tetrahedral Cu coordination (CN~=~4) is compatible with p-type TCM character. Second, the CMVB mechanism is not restricted
to oxide chemistries: I~$5p$ states play the role of the anion $p$
states in the hybridization at the VBM, demonstrating that the
framework extends naturally to halide-based systems. Third, the
energetic position of the Cu $d$ states relative to the VBM,
$\epsilon_d$, provides a quantitative descriptor for valence band
dispersion that refines the qualitative CMVB prescription into a
predictive tool. In fact, $\epsilon_d$ is strongly correlated with the width of
the topmost valence band ($r = -0.75$) and moderately correlated with
the light hole effective mass ($r = 0.53$), identifying it as a useful
pre-screening quantity for future high-throughput searches. The Pauling
electronegativity of the third element X correlates with $\epsilon_d$
($r = -0.56$), providing a chemically intuitive handle for valence band
engineering. Fourth, the tendency toward Cu undercoordination, expressed
structurally as Cu vacancy formation and Cu-deficient stoichiometries,
is an intrinsic feature of halide-based systems, arising from the low
electrostatic penalty for removing a monovalent Cu$^+$ cation from a
monovalent-anion host. This directly links the coordination environment
of Cu to natural p-type carrier generation, unifying the coordination
and dopability design rules into a single principle with no direct
counterpart in the oxide-based CMVB framework.

At the same time, the dataset has inherent limitations that must be
acknowledged. The presence of both Cu and I in all structures introduces
a systematic bias. For example, $p$-$d$ hybridization and CN~=~4 for Cu are expected
by construction, making it impossible to establish sufficiency from this
dataset alone. The absence of failed candidates, i.e., materials that are
stable and transparent but have high hole effective mass, prevents a
full statistical separation of causal factors from coincidental
correlations. More targeted studies, combining the present candidates
with a contrast set of non-TCM structures, will be needed to establish
which of the identified features are truly necessary and sufficient
conditions for p-type TCM character.

The 58 candidate materials identified here, and in particular the 20
p-type degenerate semiconductors, provide concrete targets for future
experimental synthesis and characterization. Several stand out as
particularly promising. Cu$_2$FI$_2$, unreported to our knowledge, is a
p-type degenerate semiconductor with a minimum hole effective mass of
$0.262\,m_0$ and an mHSE06 gap of 3.75~eV; its non-zincblende structure
makes it qualitatively the most distinct candidate within the halide family~\cite{Glawe_2016}. Among the layered compounds, Cu$_2$BrI
and NiCu$_2$I$_4$ show the closest structural resemblance to the hexagonal CuI monolayer recently realized in graphene encapsulation~\cite{Mustonen_2022}, which is related to the layered high-temperature $\beta$ phase of CuI~\cite{Sakuma_1988}. Crystal structures of representative candidates are provided
in Fig.~S2 of the Supporting Information. The four generalizations of
the CMVB framework provide guidance for extending the search
beyond CuI-based chemistries to other halide and chalcogenide hosts
where Cu undercoordination, tetrahedral bonding, and tunable
$d$ state energetics may be similarly exploited.

\section*{Conflicts of interest}
There are no conflicts to declare.

\section*{Acknowledgements}
This work was funded by the Deutsche Forschungsgemeinschaft (DFG)
through the research unit FOR~2857 and projects BO~4280/9-1 and
BO~4280/9-2. Computational resources were provided by the Leibniz
Supercomputing Centre (LRZ) through project pn68le on SuperMUC-NG.
M.S.\ thanks Janine George and Christina Ertural for their support with
the LOBSTER package.

\section*{Data availability}
The data that support the findings of this study are available within
this article and its supplementary material, as well as in the
{\sc Alexandria} materials database, accessible and downloadable from
\url{https://alexandria.icams.rub.de/} under the terms of the
\href{https://creativecommons.org/licenses/by/4.0/}{Creative Commons
Attribution 4.0 License}.





\bibliography{Literatur.bib} 

\end{document}